# High Resolution Asynchronous Digital Event Timer

[1] Farhat Iqbal, [2] Waheed Rathore, [3] Khawar Iqbal, [4] Glulam. Jaffer

*Abstract*—Many scientific and astronomical instruments need precise time measurement with high resolution between two or more events which is very challenging since decades. Presently a fast response of high resolution 17.1ps Digital Event Timer (DET) has been implemented using FPGA. The floor plan of FPGA is purposely arranged to get the higher resolution for timing measurement. As compared to various event timers (time to digit converters), the present design has better accuracy and higher resolution. The present DET is using the clock frequency (GHz) capable of measuring time in fraction microseconds and a unique asynchronous vernier approach to measure the time fraction of clock cycle which increases its resolution compared to Digital Event Timers. Fast response and higher resolution make it a better choice for multi KHz satellite laser ranging and other LIDAR applications.

## I. INTRODUCTION

Time-To-Digital Converters (TDCs) assume an imperative role in all computational frameworks exactingly. They may be in Phase-Locked Loops (PLLs), where they measure the distinction between the loop and the reference clock to avoid clock drift for Time of Flight (ToF) applications where the time between an emission and reception is measured to discover information about an object from which the signal was reflected or the environment through which the signal passed [1]. Likewise, there are additionally quantum adaptations of these applications, where the signal is a single quantum, and the PLL or ToF estimation should perform well, not withstanding some quanta being lost in-flight [2]. They additionally show up in medical imaging, as certain frameworks, for example, Positron Emission Tomography (PET) and Fluorescence Lifetime Imaging (FLIM) utilize the ToF or absorption time of tissues or substances to make internal image of complex structures similar to human body [3].

The device capable of measuring time difference between two pulses or events is called event timer. Event Timers have been designed by using both analog and digital techniques [4].

Analog techniques are mainly based on time measurement of capacitor charge and discharge methods. The main problems in analog measurement are long conversion time, poor stability, nonlinearity and difficulty to achieve high resolution, while the digital methods are used for large measurement and related result can be enhanced by averaging.

There are numerous ways and platforms to implement DETs, from which the resolution and performance of the measurement can be fluctuated [5]. The core objective of the present Digital Event Timer (DET) is higher resolution, whereas resolution is the minimum value of time difference which can be measured precisely. Efforts have been made to improve the Pico-second resolution through vernier, time-stretching and tapped delay lines. The present DET can be used widely in verity of scientific applications such as Time-Of-Flight (TOF) measurements, laser range finder, satellite positioning, frequency counter, nuclear physics, on-chip jitter measurements etc.

The present technique may be implemented in ASIC (Application Specific Integrated Circuits) or inside FPGA devices. However, the ASIC base devices are costly in particular if produced in small quantities and design process is complex due to the long turn-around time and layout phase [6]. On the other hand, Field-Programmable Gate Arrays (FPGA) base devices are low cost, short design time and are commercially available. FPGAs have pre-constructed grid of gates and lookup-tables (LUT) that can be configured to act like arbitrary digital circuits. It gives a flexible way of implementing a large variety of circuits [7]. Implementation of the DET requires delay cells with closely corresponding delays, as the difference in delay between latch and buffer cell determines the resolution of the DETs.

The first Digital Event Timer termed as time to digital converter implemented in FPGA was proposed by Kalisz, et al in 1997. The technique used in his research measures time difference between a latch delay and a buffer delay of Quick Logic's FPGA and achieved a time resolution of 500 ps. In 1997, Kalisz et. al, proposed an FPGA-based approach: their design used a variation of conventional delay line and offered a time resolution of 200 ps [8]. In 2000, rapid progress in electronics technology allowed them to achieve a time resolution of 100 ps [9]. Two different digital delay line circuits have been designed and tested by the authors and resolution values between 50 ps and 500 ps have been

High Resolution Asynchronous Digital Event Timer

achieved with this technology [9, 10]. In 2008, a digital event timer with fast response, medium resolution was formulated having 500 ps resolution using chain of delay line [11]. In 2015, implementation of a 30 ps resolution time to digital converter in FPGA was designed [12]. Resolution in picoseconds of an event timer is of prime importance in range finding applications[3]. For example, in optical range finder the uncertainty in measuring the event time is 0.3 mm per ps, which is directly multiplied with the resolution of timing measurement i.e. DET with 500 ps resolution will result in uncertainty of 1.5 cm. While a higher resolution event timer 17 ps will reduce the uncertainty cripplingly. Millimeter ranging accuracy and higher repetition rate are the ultimate goals in satellite laser ranging [13] and various LIDAR applications.

The present asynchronous design is implemented inside the FPGA. The basic variation between a synchronous design and asynchronous design is that the synchronous design is dependent on the clock pulse with which the system has been synchronized, and resolution is reliant on the frequency i.e. a clock of 1 MHz indicates a resolution of 1,000,000 ps. If event 1 starts in between the first clock pulse and event 2 starts just after the event 1 with minor difference. The time measurement of event 1 and event 2 will be same as shown in Figure 1.

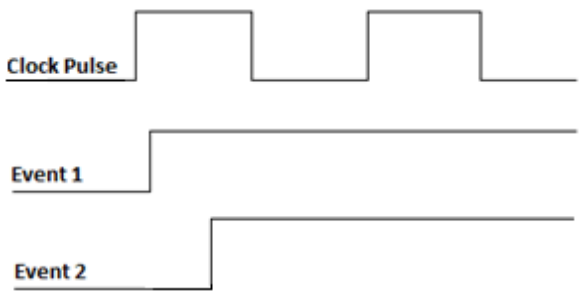

Fig 1: Error in calculating events time

While asynchronous design can get higher resolution by using event triggered measurement. It utilizes a clock to measure multiple clock cycles time and a small delay line as a vernier to measure the difference between two events within one clock cycle. To achieve maximum possible resolution with the available device, asynchronous approach is an optimal choice.

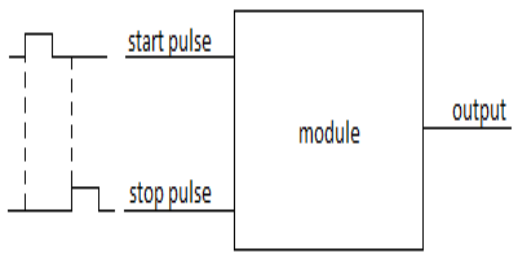

Fig 2: Event delay detection module

## II. PRINCIPLE OF OPERATION

Event timer measures the accurate time between the two instances, it records the time interval between the start pulse and stop pulse. When an event starts, the start pulse generator generates the start pulse and when the event stops the stop pulse generator generates the stop pulse. Both these pulses pass through the module and gives out the time difference result at output pin as shown in Figure 2.

The system asynchronously depends on two pulses, start pulse and stop pulse. The basic design comprises Vernier delays shown in Figure 3.

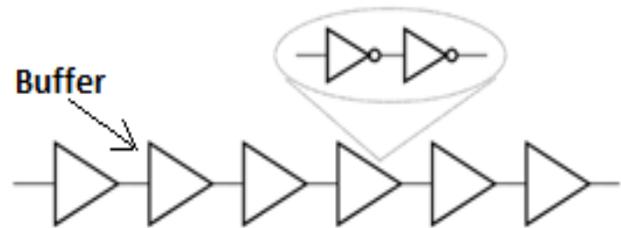

Fig 3: Basic design of buffer chain

The implementation of Vernier delays need an understanding of FPGA architecture. The architecture of an FPGA consists of Logic Cells [LCELL] and Lookup tables. For the purpose of constructing of delay lines, logic cells are used to construct a delay buffer. LCELL implements a buffer gate and a latch. Buffers are used for delay purpose and latches are used to store the buffer output. Several number of delay buffers in series are constructed to design a chain. Figure 4 shows the construction of single chain inside the

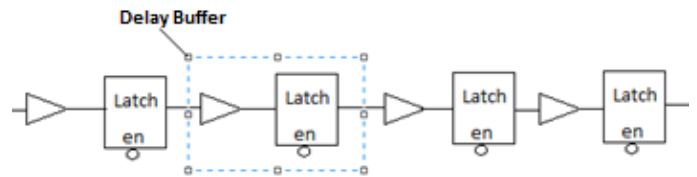

FPGA.

Fig 4: Construction of single chain inside the FPGA

Several numbers of such chains are constructed inside the FPGA having increasing number of delay buffers. Last chain consists of only one delay buffer. These delay chains are organized in descending order such that if one chain has N delay buffers and the next one has N-1 delay buffers. The resolution is difference between two consequent chains is one time constant/resolution as shown in Figure 5.

### A. Event delay detection module

When event starts, the start pulse generator generates the start pulse; it travels through all the delay chains. The start pulse is passing through each delay buffer of the chain and gives the value 0/1 as output. An exemplary circuit of four such delay chains is shown in Figure 5 and described for clarity.

The four outputs are fed as 4 inputs to four input-OR gates. The output of 'OR' gate is the input into the two input 'AND' gate. Stop pulse generator generates the stop pulse at the second input pin of 'AND' gate. The output of the 'AND' gate is further input into the two input 'OR' gate.

Second input of the 'OR' gate is the output of respective delay chain. Finally, output of the 'OR' gate is working as input of the enable pin of Latch. This completes the basic design of our event timer as shown in Figure 6.

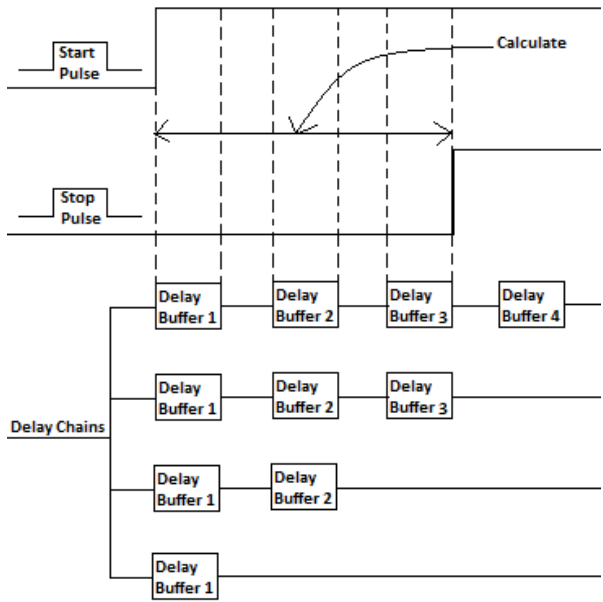

Fig 5: Digital event timer design

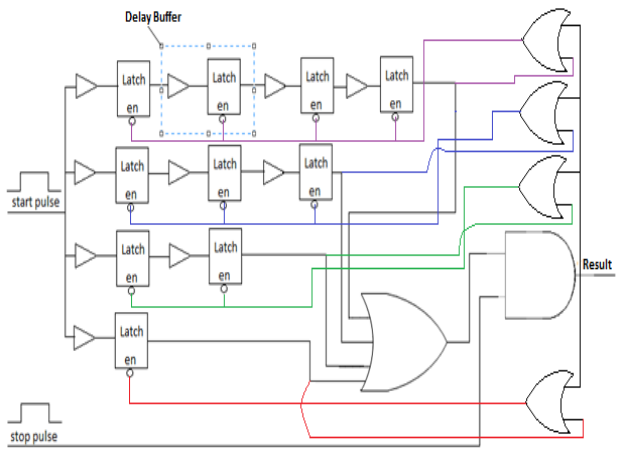

Fig 6: working of event delay module

The reset is used to reset all the values of delay buffers to be zero and latch is in the write state. When a start pulse is generated it will simultaneously pass through all the delay chains. Now we find that from which delay chain start signal is completely passed which enables us finally to calculate difference of time between two events i.e. start and stop event. The result is the multiple of delay buffer used as time constant. The Latches will freeze the present values. The delay of one delay buffer in the chain is measured to be 11.4 ps and if floor plan of FPGA is arranged manually then delay of wire which connected the two-delay buffer is assumed to be delay equal to one inverter which is 5.7 ps. A similar larger design with 10 such chains is realized inside the FPGA.

### B. Event delay calculation module

The time calculating module consists of Latches, look up table (LUT) and Multiplexers (MUX). When a signal is passed through the delay chains, latches stores 1/0 depending on if the signal has passed through or not. MUX takes the input signal as selector pin from the chains through which start pulse completely passes. LUT contains the possible outcomes and MUX generates output result. Selector pin notify the MUX, what will be the final value of output as shown in Figure 7.

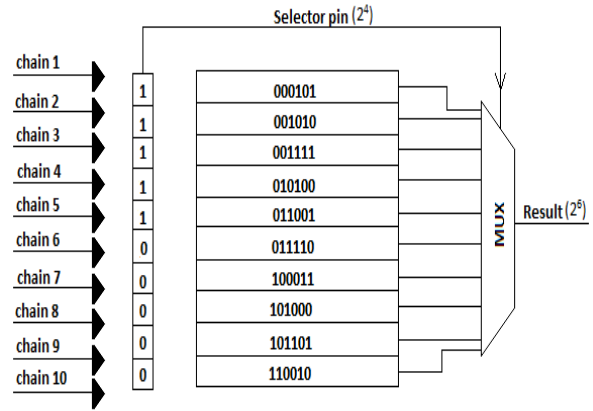

Fig 7: working of multiplexer

### III. DESIGN AND SIMULATION

#### A. Design

Code is written in Verilog language by using software Quarts II. When compiling the code, it will generate the RTL design. Two modules comp2 (Event delay detection module) and cal (Event delay calculation module) are generated as shown in Figure 8.

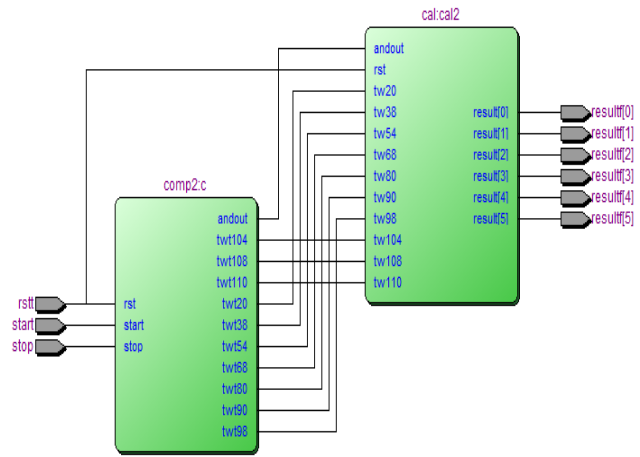

Fig 8: RTL design of event timer

#### B. Event delay detection module

When expanding the RTL of comp2 module it gives the following basic structure as shown in Figure 9, which consist of start signal, stop signal, reset pin, basic logic gates, latches and chain of buffer (inverter) connected with each other in cascading style, that generates the output result.

High Resolution Asynchronous Digital Event Timer

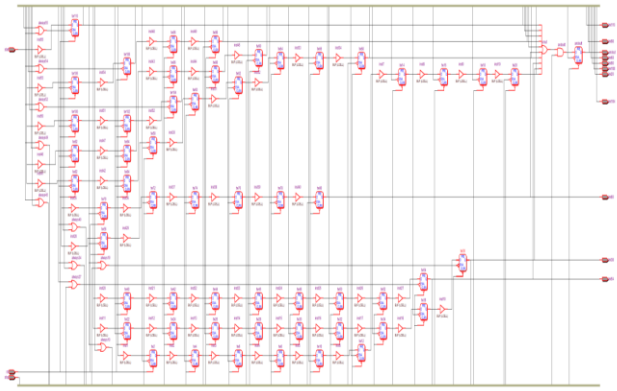

Fig 9: *RTL of* event delay detection module

C. *Event delay calculation module*

When expanding the RTL of call-module it gives the following basic structure as shown in Figure 10. The output of the comp2 module feeds as an input into the second cal module that gives the final calculation of the event occurring times in the decimal number of the event timer.

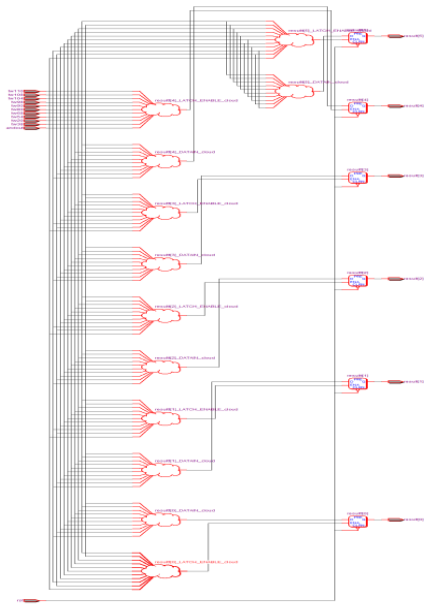

Fig: 10.   RTL of event delay calculation module

## IV.   TEST BENCH

A. *Measuring time interval for 10 ps*

The simulation of Verilog code is done by using Model-Sim that will generate the waves shown in Figure 11.The system is negative enabled. Firstly, '**reset**' the event timer to remove the garbage values. The start signal starts at 10 ps when event start. Then signal passing through the chains in ascending order. After completion of event at 20 ps, stop signal will be generated and gives the output at '**andout**' at 20 ps. So, the result will be stop signal minus start signal that comes out to be 10 ps. The results acquired for different number of LCELLs' is tabulated in Table I.

TABLE I
Event Delay Calculation Results

| No. Of LCELL | Delay per interconnect [Ps] | Delay/ LCELL (chain)[Ps] | Total delays [Ps] | Difference in two delays [Ps] |
|---|---|---|---|---|
| 1 | 5.7 | 11.4 | 17.1 | 17.1 |
| 2 | 11.4 | 22.8 | 34.2 | 17.1 |
| 3 | 17.1 | 34.2 | 51.3 | 17.1 |
| 4 | 22.8 | 45.6 | 68.4 | 17.1 |
| 5 | 28.5 | 57 | 85.5 | 17.1 |
| 6 | 34.2 | 68.4 | 102.6 | 17.1 |
| 7 | 39.9 | 79.8 | 119.7 | 17.1 |
| 8 | 45.6 | 91.2 | 136.8 | 17.1 |
| 9 | 51.3 | 102.6 | 153.9 | 17.1 |
| 10 | 57 | 114 | 171 | 17.1 |

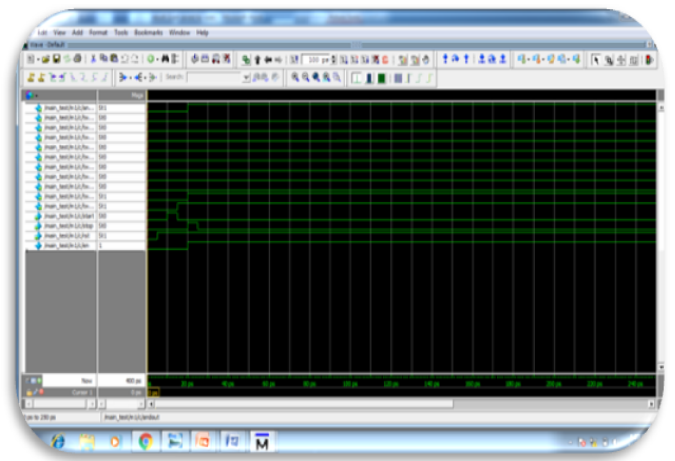

Fig 11: Simulation of signal passing at 10 ps

B. *Calculation for 10 ps*

The calculation will be done on the final signal coming from the '**andout**' and gives the result as '**result**' in decimal form (10 ps) as shown in the Figure 12.

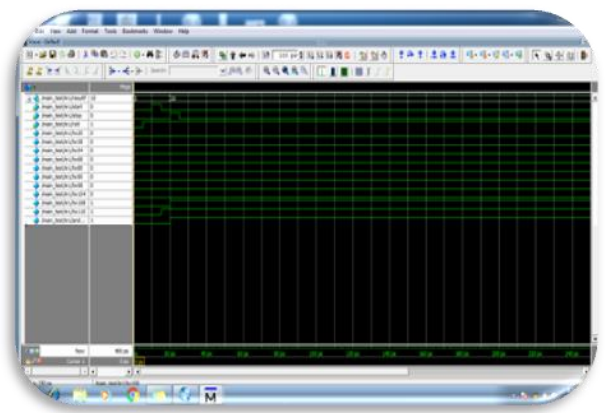

Fig 12: Calculation at 10 ps simulation

## VI. TEST RESULTS

The chain of delay line consists of double inverters as delay elements (LCELL) shown in Fig 13. The delay of the single inverter in 65 nm technology which is used by cyclone III is 5.7 ps. So single buffer takes the 11.4 ps delay time.

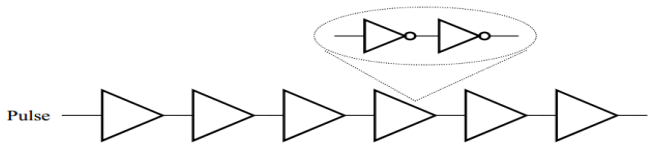

Fig 13: Delay line (LCELL)

If we assume the two-buffer place next to each other by using manual floor planning, then interconnects delay will be assumed as delay of one inverter. So, the final delay of buffer and interconnects will be 17.1 ps.

The final calculation of events that held at different time interval is shown in Table I and a Fujitsu report depicting speed performance improvements is placed as Figure 14.

|  | 65nm CS200 (ps/gate) | 90nm CS100 (ps/gate) | Delay Improvement |
|---|---|---|---|
| Inverter | 5.7 | 7.0 | 19% |
| 2-input NAND | 8.7 | 11.4 | 24% |

Fig 14: Fujitsu Report

## V. CONCLUSION

The fast response digital event timer is implemented based on above mentioned vernier method. The vernier uses standard buffer gates. To implement a series of identical, cascaded delay gates we had to disable the automatic optimization of FPGA device. The event pulse starts traveling through several parallel chains of Gates; each chain consists of an increasing number of gates; the following clock pulse is used as a 'STOP' pulse for this simple vernier. If the start pulse reaches the end of a chain before the 'STOP' pulse, a "1" is latched into an output register; if not: a "0". The bits in this output register represent a measure for time interpolation; together with the latched actual reading of the coarse clock this forms a 17 ps resolution event time.

The present approach for TDCs utilizes a mixture of both synchronous and asynchronous digitizing of time interval between two input signals, that enhances the accuracy of time measurement. The proposed design also improves the circuit simplicity, dynamic range, resolution, and chip area in comparison with the previous works.

Results show that present design is a cost-effective solution for high resolution 17 ps event timer. Resolution wise this technique is placed between asynchronous and synchronous design if used along with synchronous counter to count the pulses. The proposed asynchronous part of design may be used separately as vernier to measure the additional delay less than a clock pulse.

The fundamental advantages and architecture choices enabled by the use of the TDC are being extended to faster LIDAR applications [14] require further research and higher resolution timing devices in near future. The advent of new, faster and latest minimized semiconductor devices may lead to even higher resolution digital counters and short delay elements resulting in better resolution in future.